\documentclass[a4paper, 11pt]{article}
\usepackage{amsfonts}
\usepackage{dsfont}
\usepackage{mathrsfs}
\usepackage{amssymb}
\usepackage{bbm}
\usepackage{epsfig}
\usepackage{amsmath}
\usepackage{appendix}
\usepackage{float}
\usepackage[english]{babel}

\def\Reals{\mathop{\hbox{\mit I\kern-.2em R}}\nolimits}

\def\Complexes{{\hbox{\mit C\kern-.46em
            \vrule depth 0ex height 1.4ex width .05em\kern.41em}}}

\newtheorem{thm}{Theorem}
\newtheorem{defn}{Definition}

\newtheorem{lem}{Lemma}
\newtheorem{remark}{Remark}
\newtheorem{prop}{Proposition}

\title{ \bf The Role of Persistent Graphs\\ in the Agreement Seeking of Social Networks \footnote{This work has been supported in part
by the Knut and Alice Wallenberg Foundation, the Swedish Research
Council and  KTH SRA TNG.}}
\date{}

\author{Guodong Shi and Karl Henrik Johansson\thanks{G. Shi and K. H. Johansson are with ACCESS Linnaeus Centre, School of Electrical Engineering,
Royal Institute of Technology, Stockholm 10044, Sweden.
       Email: {\tt\small guodongs@kth.se, kallej@kth.se}}
}
\begin{document}
\maketitle
\begin{abstract}
This paper investigates the role persistent arcs play for a social network to reach a global belief agreement under discrete-time or continuous-time evolution. Each (directed) arc in the underlying communication graph is assumed to be associated with a time-dependent weight function which describes the strength of the information flow from one node to another.  An arc is said to be persistent if its weight function has infinite $\mathscr{L}_1$ or $\ell_1$ norm for continuous-time or discrete-time belief evolutions, respectively.   The graph that consists of all persistent arcs is called the persistent graph of the underlying network.  Three  necessary and sufficient conditions on agreement or $\epsilon$-agreement are established, by which we prove that the persistent graph fully determines the convergence to a common opinion in social networks. It is shown how the convergence rates explicitly depend on the diameter of the persistent graph.  The results adds to the understanding of the fundamentals behind  global agreements, as it is only persistent arcs that contribute to the convergence.
\end{abstract}

{\bf Keywords:}  Consensus, Persistent Graphs, Social Networks, Dynamical Systems

\section{Introduction}
Recent years have witnessed wide research interest in opinion dynamics and information aggregation of social networks.  Individuals are equipped with {\it beliefs} or {\it opinions} which updated as  information is exchanged from time to time; how beliefs are propagated depends on the interactions between individuals.

DeGroot's model is a classical  model on belief evolution \cite{degroot}. It is simply formulated as a  discrete-time linear system, where the state transition matrix is time-invariant and row stochastic. The $ij$-entry of the transition matrix of DeGroot's model represents the {\it weight} of the arc which marks the influence of $j$ to $i$. The convergence to an agreement  is equivalent  to the convergence to a stationary distribution of the finite-state Markov chain given by the same transition matrix. Results from Markov chain analysis can therefore be used in the agreement analysis \cite{degroot,wolf,haj}.   Variations of DeGroot's model are considered in \cite{social1, golub, daron, como} for the study of opinion dynamics in social networks. Here if an asymptotic belief agreement can be reached or not has always been a central question.

Consensus problems which are very related to DeGroot's model appear in many different contexts in the study of computer science and engineering,  e.g., decentralized and parallel computations \cite{tsi,cs2,cs3}, coordinations of autonomous agents \cite{jad03, mar, mor, ren05} and sensor networks \cite{sn1,sn2,sn3}. Agreement seeking has been extensively studied in the literature for both discrete-time and continuous-time models \cite{jad03,tsi,tsi2,mor,boyd,caoming1,caoming2, fax, saber04, lin07,shi09,lwang}.

The communication graph plays an important role in  proper conditions to ensure a consensus. In most existing work, the arc weights, which reflect the strength of the influence from one node to another, are assumed to either be constant whenever two nodes meet with each other \cite{degroot,boyd,saber04}, or in a compact set with positive lower and upper bounds \cite{tsi2,jad03,caoming1,caoming2,daron}. However, in reality, the arc weights may vary in a wide range, and may even fade away. Moreover, different arcs may have quite different persistency properties.  For instance, the opinion of people may be heavily influenced over short time periods by political campaigns, but over long time periods persistent links to family and friends might be more important.  This is to say, the weights of the opinions from different sources are in practice  generally time-varying and highly irregular  over the underlying communication graph, and therefore, links can be impulsive, vanishing, persistent, etc. Then an interesting question arises: are  there certain arcs which are the ones that actually generate the convergence to a consensus and how do their graph properties influence the convergence rate?

The central aim of the paper is to build a model to classify different arcs in the underlying communication graph, and then give a precise description on how the persistent  arcs indeed determine the agreement seeking. We define the persistent graph as the graph having links whose weight functions have infinite $\mathscr{L}_1$ or $\ell_1$ norm for continuous-time or discrete-time belief dynamics, respectively. Global agreement and $\epsilon$-agreement are defined as whether the maximum state difference converges to zero, and whether the convergence is exponentially fast, respectively. For the discrete-time case, a  necessary and sufficient condition is obtained on $\epsilon$-agreement under general stochasticity, self-confidence and arc balance assumptions. Then for the continuous-time case, two necessary and sufficient conditions are established on global agreement and $\epsilon$-agreement, respectively. In this way, we precisely state how the persistent graph plays a fundamental role in consensus seeking. Additionally, comparisons of our new conditions are given with existing results and the relations between the discrete-time and continuous-time evolutions are highlighted.

The rest of the paper is organized as follows. In Section 2, we introduce the network model and define the problem of interest. Then in Sections 3 and 4, the main results and convergence analysis are presented for discrete-time and continuous-time dynamics, respectively. Finally some discussions and concluding remarks are given in Sections 5 and 6.

\section{Problem Definition}
In this section, we present the social network model and define the considered problem. To this end, we first introduce some basic  graph theory \cite{god}.

A (simple) {\it digraph} $\mathcal
{G}=(\mathcal {V}, \mathcal {E})$ consists of a finite set
$\mathcal{V}=\{1,\dots,n\}$ of nodes and an arc set
$\mathcal {E}$, where  each {\it arc}  $(i,j)\in\mathcal {E}$ is an ordered pair
 from node $i\in \mathcal{V}$ to another  node $j\in\mathcal{V}$. If the
arcs are pairwise distinct in an alternating sequence
$ v_{0}e_{1}v_1e_{2}v_{2}\dots e_{k}v_{k}$ of nodes $v_{i}$ and
arcs $e_{i}=(v_{i-1},v_{i})\in\mathcal {E}$ for $i=1,2,\dots,k$,
the sequence  is called a (directed) {\it  path} with {\it length} $k$.
A path from $i$ to
$j$ is denoted $i \rightarrow j$, and the length of $i \rightarrow j$ is denoted  $|i \rightarrow j|$. A path with no repeated nodes is called a {\it simple} path. If there exists a path from node $i$ to node $j$,
then node $j$ is said to be reachable from node $i$.  Each node is thought to be reachable by itself. A node $v$ from which any other node is
reachable  is called a {\it center} (or a {\it root}) of $\mathcal {G}$.  $\mathcal
{G}$ is said to be {\it strongly connected}  if it contains path $i \rightarrow j$ and $j \rightarrow i$ for every pair of nodes $i$ and $j$; $\mathcal {G}$ is said to be {\it
quasi-strongly connected} if $\mathcal {G}$ has a center
\cite{ber, lin07}.

The {\it distance} from $i$ to $j$, $d(i,j)$,   is defined as the length of a shortest (simple) path $i \rightarrow j$ when $j$ is reachable from $i$, and the {\it diameter} of $\mathcal
{G}$ as $d_0=\max\{d(i,j)|i,j \in\mathcal
{V},\ j\mbox{ is reachable from}\ i\}$.

\vspace{2mm}

In this paper, we consider a social network model  with node set $\mathcal{V}=\{1,\dots,n\}$.  Let the digraph $\mathcal {G}_\ast=(\mathcal{V},\mathcal{E}_\ast)$ denote the {\it underlying}  graph of the considered social network. The underlying graph indicates all potential interactions between nodes. Node $j$ is said to be a {\it neighbor} of $i$ at time $t$ when there is an arc $(j,i)\in \mathcal
{E}_{\ast}$; each node is supposed to be a neighbor of itself. Let $\mathcal{N}_i=\{i\}\cup\{j:(j,i)\in\mathcal{E}_\ast\}$ denote the neighbor set of node $i$.

Let $x_i(t)\in\mathds{R}$ be the {\it belief} of node $i$ at time $t$. Time is either discrete or continuous. The initial time is $t_0\geq0$ in both cases and each node is equipped with an initial belief $x_{i}(t_0)$. The belief updating rule is in  discrete time:
\begin{align}\label{9}
{x}_i(t+1)=\sum_{j\in \mathcal{N}_i}W_{ij}(t)x_j(t),\ \ i=1,\dots,n
\end{align}
and in continuous time:
\begin{align}\label{0}
\dot{x}_i(t)=\sum_{j\in \mathcal{N}_i}W_{ij}(t)\big[x_j(t)-x_i(t)\big], \ \ i=1,\dots,n.
\end{align}
Here $W_{ij}(t):[0,\infty)\rightarrow [0,\infty)$ is a nonnegative scalar function which represents the weight of arc $(j,i)$. Clearly $W_{ij}(t)$ describes the strength of the influence of node $j$ on $i$. Since $W_{ij}(t)=0$ may happen from time to time, the graph is indeed time-varying.

We define
$$
\psi(t)\doteq\min_{i\in\mathcal {V}}\{x_i(t)\} ,\quad \Psi(t)\doteq\max_{i\in\mathcal {V}} \{x_i(t)\}
$$
as the  minimum and maximum state value at time $t$, respectively. Then
$$
\mathcal {H}(t)\doteq\Psi(t)-\psi(t)
$$
 is a natural  agreement measure marking the maximum distances between the individual beliefs. The  considered global agreement  and $\epsilon$-agreement  for both the discrete-time and continuous-time updating rules are defined as follows.

\begin{defn}
(a) Global {\it agreement}  is achieved if for any $x(t_0)\doteq(x_1(t_0) \dots x_n(t_0))^T\in \mathds{R}^{n}$, we have
\begin{equation}
\lim_{t\rightarrow \infty} \mathcal {H}(t)=0.
\end{equation}

(b)  Global  $\epsilon$-agreement  is achieved if there exist two constants $0<\epsilon<1$ and $T_0>0$ such that for any  $x(t_0)\in \mathds{R}^{n}$ and $t\geq t_0$, we have
\begin{equation}
\mathcal {H}(t+T_0)\leq \epsilon \mathcal {H}(t).
\end{equation}
\end{defn}
\begin{remark}
A global agreement only requires that $\mathcal {H}(t)$ will converge to zero as $t$ tends to infinity. If it is further required that the convergence speed is at least exponentially fast,  we use global $\epsilon$-agreement. This definition of $\epsilon$-agreement and other similar concepts have been widely used to characterize the convergence rate of consensus evolutions in the literature, e.g., \cite{boyd, tsi2, olshevsky1, olshevsky}.
\end{remark}

The goal of this paper is to distinguish the arcs from the underlying graph that are  {\it persistent} over a long time range and how they influence global agreement.  To be precise, we impose the following definition for persistent arcs and persistent graphs based on the $\mathscr{L}_1$ or $\ell_1$ norms of the weight functions.
\begin{figure}[H]
\centerline{\epsfig{figure=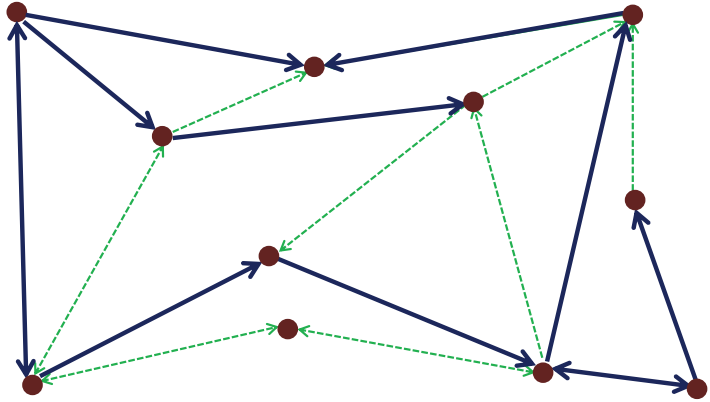, width=0.60\linewidth=0.25}}
\caption{The underlying graph consists of persistent arcs (solid) and vanishing arcs (dashed). The persistent graph is shown to play a fundamental role for the convergence to an agreement. }\label{sss}
\end{figure}
\begin{defn}
(a) An arc $(j,i)\in \mathcal{G}_\ast$ is  a persistent  arc of the discrete-time updating rule (\ref{9})  if
$$
\sum_{t=0}^\infty W_{ij}(t)=\infty,
$$
and a persistent  arc of the continuous-time updating rule  (\ref{0}) if
$$
\int_{s}^\infty W_{ij}(t)dt=\infty\ \mbox{for all} \ s\geq 0.
$$

(b) The graph $\mathcal{G}^p=(\mathcal{V},\mathcal{E}^p)$ that consists of all persistent arcs is called  the persistent graph.
\end{defn}

Next, in Sections 3 and 4, we will investigate the discrete-time and continuous-time updating rules, respectively.  We will establish sufficient and necessary conditions on global agreement and $\epsilon$-agreement, which illustrate that the notion of persistent graphs is critical to the convergence.

\section{Discrete-time Belief Evolution}
In this section, we focus on the discrete-time belief evolution (\ref{9}). In order to obtain the main result, we need the following assumptions.

{\bf A1} {\it (Stochasticity)} $\sum_{j\in \mathcal{N}_i}W_{ij}(t)=1$ for all $i\in\mathcal{V}$ and $t\geq 0$.

{\bf A2} {\it (Self-confidence)} There exists $0<\eta<1$ such that $W_{ii}(t)\geq \eta$ for  $i\in\mathcal{V}$ and $t\geq 0$.

{\bf A3} {\it (Arc Balance)} There exists a constant $A>1$ such that for any two arcs $(j,i), (m,k)\in\mathcal{E}^p$ and $t\geq 0$, we have
$$
A^{-1} W_{ij}(t)  \leq  W_{km}(t) \leq A W_{ij}(t).
$$

The main result for the discrete-time updating rule (\ref{9}) on global $\epsilon$-agreement is as follows.

\begin{thm}\label{thm0}
Suppose A1, A2 and A3 hold. Global $\epsilon$-agreement is achieved for (\ref{9}) if and only if

 (a) $\mathcal{G}^p$ is quasi-strongly connected;

 (b) there exist a constant $a_\ast>0$ and an integer $T_\ast>0$ such that $\sum_{s=t}^{t+T_\ast-1} W_{ij}(s)\geq a_\ast$ for all $t\geq0$ and $(j,i)\in \mathcal{E}^p$.

In fact, if (a) and (b) hold, then we have
 \begin{align}
\mathcal{H}(t+d_0T_\ast)\leq\Big(1- \frac{\eta^{d_0T_\ast}}{2}\cdot \big(\frac{a_\ast}{T_\ast}\big)^{d_0}\Big) \mathcal{H}(t)
\end{align}
for all $t\geq t_0$, where $d_0$ represents the diameter of $\mathcal{G}^p$.
\end{thm}

\begin{remark}
Consensus convergence for many variations of (\ref{9}) has been extensively studied in the literature, e.g., \cite{bert,daron,golub,wolf,haj,jad03,ren05,mor,caoming1}. As for convergence rate, a relatively conservative bound is given in \cite{bert,jad03}, and then  generalized in \cite{caoming2, olshevsky1}. Recently a sharper bound for convergence rate was obtained  in \cite{olshevsky}. The self-confidence condition A2 is generally not necessary to ensure a consensus, but the convergence properties may be quite different  without A2, especially for the case with time-varying graphs.
 \end{remark}

 \begin{remark}
 Most of existing results are based on the assumption that all weight functions $W_{ij}(t)$ in the underlying graph have a positive lower bound whenever they are not zero. Here we just need the self-loop weights, $W_{ii}(t),i=1,\dots,n$, to have a positive lower bound. As indicated by the proof below,  the sufficiency statement of Theorem \ref{thm0} relies on the self-confidence assumption A2, while the arc balance assumption A3 is used in the necessity part.
\end{remark}

Before we state the proof,  we introduce some more notations, which will be used throughout the rest of the paper. For two sets $S_1$ and $S_2$, $S_1\setminus S_2$ is defined as $S_1\setminus S_2=\{z:z\in S_1, z\notin S_2\}$. For the underlying graph $\mathcal{G}_\ast=(\mathcal{V}, \mathcal{E}_\ast)$ and the persistent graph $\mathcal{G}^p=(\mathcal{V}, \mathcal{E}^p)$, we denote
\begin{align}\label{notation3}
\theta(t)=\sum_{(j,i)\in \mathcal{E}_\ast\setminus\mathcal{E}^p}W_{ij}(t), \quad
\end{align}
and
\begin{align}\label{notation1}
\xi^+(t;m)=\sum_{j\in\mathcal{N}_m\setminus \{m\}}W_{mj}(t), \quad {\xi}_0^+(t;m)=\sum_{j\in\mathcal{N}_m\setminus \{m\}, (j,m)\in \mathcal{E}^p}W_{mj}(t).
\end{align}

In the following two subsections, we prove the necessity and sufficiency parts of Theorem \ref{thm0}, respectively.

\subsection{Necessity}
We need to show that a global $\epsilon$-agreement cannot be achieved without either condition $(a)$ or $(b)$.

The upcoming analysis  relies on the following well-known lemmas.
\begin{lem}\label{lemd1}
Suppose $0\leq p_k<1$ for all $k$. Then $\sum_{k=0}^\infty p_k=\infty$ if and only if $\prod_{k=0}^{\infty}(1-p_k)=0$.
\end{lem}
\begin{lem}\label{lemd2}
$\log(1-t)\geq -2t$ for all $0\leq t\leq {1}/{2}$.
\end{lem}

We have the following proposition indicating that $\mathcal{G}^p$ being quasi-strongly connected is not only a necessary condition for  (\ref{9}) to reach global $\epsilon$-agreement, but also necessary for (simple) global agreement, even in the absence of assumptions A2 and A3.

\begin{prop}\label{prop1}
Suppose A1 holds. If  global agreement is achieved for (\ref{9}), then $\mathcal{G}^p$ is quasi-strongly connected.
\end{prop}
{\it Proof.} Suppose $\mathcal{G}^p$ is not quasi-strongly connected. Then there exist two distinct nodes $u$ and $w$ such that $\mathcal {V}_u\cap \mathcal {V}_w=\emptyset$, where $\mathcal {V}_u=\{\mbox{nodes\ from\ which\ $u$\ is\ reachable\ in}\  \mathcal
{G}^p\}$ and  $\mathcal {V}_w=\{\mbox{nodes\ from\ which\ $w$\ is\ reachable\ in}$\ $\mathcal
{G}^p\}$. Moreover, there is no arc entering either $\mathcal{V}_u$ or $\mathcal{V}_w$ in the persistent graph $\mathcal
{G}^p$.  Let $x_{i}(t_0)=0$ for all $i\in\mathcal{V}_u$, and $x_i(t_0)=1$ for all $i\in\mathcal{V}\setminus \mathcal{V}_u$. Denote $\ell(t)=\max_{i\in\mathcal{V}_u} x_i(t)$ and $\hbar(t)=\min_{i\in \mathcal {V}_w}x_i(t)$. We define $g^+(t;m)=\sum_{j\in\mathcal{N}_m,j\notin \mathcal{V}_u}W_{mj}(t)$ for $m\in\mathcal{V}_u$ and $f^+(t;k)=\sum_{j\in\mathcal{N}_k,j\notin \mathcal{V}_w}W_{kj}(t)$ for $k\in\mathcal{V}_w$. We further denote
$$
\zeta_u^+(t)=\sum_{m\in\mathcal {V}_u }g^+(t;m);\quad \zeta_w^+(t)=\sum_{k\in\mathcal {V}_w }f^+(t;k).
$$

It is straightforward to see that $\psi(t)$ is non-decreasing and $\Psi(t)$ is non-increasing for (\ref{9}). It follows that $x_i(t)\in[0,1]$ for all $i$ and $t\geq t_0$. There are two cases.

\begin{itemize}
\item[(i).]  First, for any $m\in\mathcal{V}_u$, we have
\begin{align}
x_{m}(t_0+1)=\sum_{j\in\mathcal{N}_m}W_{mj}(t_0)x_j(t_0)\leq 0\cdot\big(1- g^+(t_0;m)\big)+1\cdot g^+(t_0;m)\leq \zeta_u^+(t_0),\nonumber
\end{align}
which yields $\ell(t_0+1)\leq \zeta_u^+(t_0)$ immediately. Then, for the next slot we have that for any $m\in\mathcal{V}_u$,
\begin{align}
x_{m}(t_0+2)&=\sum_{j\in\mathcal{N}_m}W_{mj}(t_0+1)x_j(t_0+1)\nonumber\\
&\leq\sum_{j\in\mathcal{N}_m,j\in \mathcal{V}_u}W_{mj}(t_0+1) \ell(t_0+1)+ \sum_{j\in\mathcal{N}_m,j\notin \mathcal{V}_u}W_{mj}(t_0+1)\cdot1\nonumber\\
&= \zeta_u^+(t_0)\cdot\big(1- g^+(t_0+1;m)\big)+ g^+(t_0+1;m)\nonumber\\
&\leq \zeta_u^+(t_0)+\zeta_u^+(t_0+1),
\end{align}
which leads to $\ell(t_0+1)\leq \zeta_u^+(t_0)+\zeta_u^+(t_0+1)$.
Continuing we get that for any $s=1,2,\dots$, we have
\begin{align}\label{d100}
\ell(t_0+s)\leq \sum_{t=t_0}^{t_0+s-1} \zeta_u^+(t)\leq \sum_{j=t_0}^\infty \theta(t)<\infty
\end{align}
because there is no arc entering $\mathcal{V}_u$ in the persistent graph $\mathcal
{G}^p$.

\item[(ii).] Consider now $\mathcal{V}_w$. According  to the definition of $\theta(t)$, there exists $T_1>0$ such that when $\theta(t)<1, t\geq T_1$. Let $t_0\geq T_1$. Then we have $\zeta_w^+(t)\leq \theta(t)<1$ for all $t\geq t_0$ since there is no arc entering $\mathcal{V}_w$ in the persistent graph $\mathcal
{G}^p$.

Similarly we  obtain $\hbar(t_0+1)\geq  1-  \zeta_w^+(t_0)$ since for any $k\in \mathcal{V}_w$, we have
\begin{align}
x_{k}(t_0+1)=\sum_{j\in\mathcal{N}_k}W_{kj}(t_0)x_j(t_0)\geq  0\cdot f^+(t_0;k)+1\cdot \big(1- f^+(t_0;k)\big)\geq 1-  \zeta_w^+(t_0).\nonumber
\end{align}
Furthermore,  for any $k\in \mathcal{V}_w$, one has
\begin{align}
x_{k}(t_0+2)&\geq 0\cdot f^+(t_0+1;k)+\big(1- f^+(t_0;k)\big)\cdot \big(1-\zeta_w^+(t_0)\big)\nonumber\\
&\geq \big(1- \zeta_w^+(t_0+1)\big)\cdot \big(1-\zeta_w^+(t_0)\big),
\end{align}
and thus $
\hbar(t_0+2)\geq \big(1- \zeta_w^+(t_0+1)\big)\cdot \big(1-\zeta_w^+(t_0)\big).$
Proceeding the analysis we know   that for any $s=1,2,\dots$,
\begin{align}\label{d2}
\hbar(t_0+s)\geq \prod_{t=t_0}^{t_0+s-1} \big(1- \zeta_w^+(t)\big)\geq \prod_{t=t_0}^{\infty} \big(1- \theta(t)\big)\geq \prod_{t=T_1}^{\infty} \big(1- \theta(t)\big)\doteq \sigma_\ast>0,
\end{align}
where $\sigma_\ast$ exists from Lemma \ref{lemd1} and the definition of $\theta(t)$.
\end{itemize}

Because $\sum_{j=0}^\infty \theta(t)<\infty$, we can always choose $t_0$ sufficiently large so that $\sum_{j=t_0}^\infty \theta(t)\leq\sigma_\ast/2$. Therefore, (\ref{d100}) and (\ref{d2}) lead to $\mathcal{H}(t_0+s)\geq \hbar(t_0+s)-\ell(t_0+s)\geq \sigma_\ast/2>0$. A global agreement is thus impossible. This completes the proof. \hfill$\square$

We establish a lemma on the upper and lower bounds for some particular nodes.

\begin{lem}\label{lemd3}
Suppose A1 holds. Let $x_m(t)=\mu\psi(t)+(1-\mu)\Psi(t)$ with $0\leq \mu\leq1$. Then for any integer $T>0$, we have:
\begin{align}\label{31}
x_m(t+T)\leq  \mu\prod_{s=t}^{t+T-1}\big(1- \xi^+(s;m)\big)\cdot\psi(t)+\Big(1-\mu\prod_{s=t}^{t+T-1}\big(1- \xi^+(s;m)\big)\Big)\cdot\Psi(t).
\end{align}
and \begin{align}\label{32}
x_m(t+T)\geq  \mu\prod_{s=t}^{t+T-1}\big(1- \xi^+(s;m)\big)\cdot\Psi(t)+\Big(1-\mu\prod_{s=t}^{t+T-1}\big(1- \xi^+(s;m)\big)\Big)\cdot\psi(t).
\end{align}
\end{lem}
{\it Proof.} When $x_m(t)=\mu\psi(t)+(1-\mu)\Psi(t)$,  for time $t+1$, we have
\begin{align}\label{30}
x_m(t+1)&=\sum_{j\in\mathcal{N}_m}W_{mj}(t)x_j(t)\nonumber\\
&\leq \big(1- \xi^+(t;m)\big)\cdot\big(\mu\psi(t)+(1-\mu)\Psi(t)\big)+\xi^+(t;m)\Psi(t)\nonumber\\
&= \mu\big(1- \xi^+(t;m)\big)\cdot\psi(t)+\Big(1-\mu\big(1- \xi^+(t;m)\big)\Big)\Psi(t).
\end{align}
For time  $t+2$, we  obtain
\begin{align}
x_m(t+2)&\leq \big(1- \xi^+(t+1;m)\big)\cdot \Big[ \mu\big(1- \xi^+(t;m)\big)\cdot\psi(t)+\Big(1-\mu\big(1- \xi^+(t;m)\big)\Big)\Psi(t)\Big]\nonumber\\
&\ \ \ \ \ \ \ \ \  +\xi^+(t+1;m)\Psi(t)\nonumber\\
&= \mu\prod_{s=t}^{t+1}\big(1- \xi^+(s;m)\big)\cdot\psi(t)+\Big(1-\mu\prod_{s=t}^{t+1}\big(1- \xi^+(s;m)\big)\Big)\cdot\Psi(t).
\end{align}
Continuing, we obtain (\ref{31}).

In equality (\ref{32}) can be easily obtained using a symmetric analysis as for (\ref{31}).  \hfill$\square$

We are now in a place to present the following conclusion, which shows  the necessity of condition $(b)$ in Theorem \ref{thm0}.
\begin{prop}\label{prop2}
Suppose A1 and A3 hold. If  global $\epsilon$-agreement is achieved for  (\ref{9}), then  there exist a constant $a_\ast>0$ and an integer $T_\ast>0$ such that $\sum_{s=t}^{t+T_\ast} W_{ij}(s)\geq a_\ast$ for all $t\geq0$ and $(j,i)\in \mathcal{G}^p$.
\end{prop}
{\it Proof.} We  prove the conclusion by contradiction. Suppose the condition does not hold. Then $\forall 0<\epsilon<1, T>0, \exists t_\ast(T,\epsilon)\geq 0$ and $(j_\ast,i_\ast)\in\mathcal{E}^p$ such that
\begin{align}\label{33}
\sum_{s=t_\ast}^{t_\ast+T-1} W_{i_\ast j_\ast}(s)<  \frac{1}{2}A^{-1}(n-1)^{-1}\cdot\log \Big(\frac{1+\epsilon}{2}\Big)^{-1}.
\end{align}
 Since $(j_\ast,i_\ast)\in\mathcal{G}^p$, it is straightforward to see that $t_\ast(T,\epsilon)\rightarrow \infty$ as $T\rightarrow\infty$ for any fixed $\epsilon$. Thus,  we can assume that (\ref{33}) also holds for the arcs in $\mathcal{E}_\ast \setminus\mathcal{E}^p$. Moreover, without loss of generality, we can also assume that $\xi^+(s;i)\leq 1/2$ for all $i$ and $t_\ast\leq s\leq t_\ast+T-1$.  With arc balance assumption A3 and Lemma \ref{lemd2}, (\ref{33}) implies
\begin{align}\label{36}
\prod_{s=t_\ast}^{t_\ast+T-1}\big(1- \xi^+(s;i)\big)=e^{\sum_{s=t_\ast}^{t_\ast+T-1}\log \big(1- \xi^+(s;i)\big)}\geq e^{-2\sum_{s=t_\ast}^{t_\ast+T-1}\xi^+(s;i)} >e^{-\log \Big(\frac{1+\epsilon}{2}\Big)^{-1}}=\frac{1+\epsilon}{2}
\end{align}
for all $i\in\mathcal{V}$.

Moreover, taking $x_m(t_\ast)=\psi(t_\ast)$ and $x_k(t_\ast)=\Psi(t_\ast)$, we know from Lemma \ref{lemd3} that
\begin{align}\label{34}
x_m(t_\ast+T)\leq  \prod_{s=t_\ast}^{t_\ast+T-1}\big(1- \xi^+(s;m)\big)\cdot\psi(t_\ast)+\Big(1-\prod_{s=t_\ast}^{t_\ast+T-1}\big(1- \xi^+(s;m)\big)\Big)\cdot\Psi(t_\ast)
\end{align}
and \begin{align}\label{35}
x_k(t_\ast+T)\geq  \prod_{s=t_\ast}^{t_\ast+T-1}\big(1- \xi^+(s;k)\big)\cdot\Psi(t_\ast)+\Big(1-\prod_{s=t_\ast}^{t_\ast+T-1}\big(1- \xi^+(s;k)\big)\Big)\cdot\psi(t_\ast).
\end{align}

Therefore, with (\ref{36}), (\ref{34}) and (\ref{35}), we eventually obtain
\begin{align}
\mathcal{H}(t_\ast+T)&\geq x_k(t_\ast+T)-x_m(t_\ast+T)\nonumber\\
&\geq \Big[ \prod_{s=t_\ast}^{t_\ast+T-1}\big(1- \xi^+(s;k)\big)+\prod_{s=t_\ast}^{t_\ast+T-1}\big(1- \xi^+(s;m)\big)-1\Big]\cdot\mathcal{H}(t_\ast)\nonumber\\
&>\big(2\cdot \frac{1+\epsilon}{2}-1\big)\mathcal{H}(t_\ast)\nonumber\\
&=\epsilon\mathcal{H}(t_\ast).
\end{align}
The desired conclusion thus follows. \hfill$\square$

The necessity claim in Theorem \ref{thm0} follows from Propositions \ref{prop1} and \ref{prop2}.

\subsection{Sufficiency}
%In order to prove the sufficiency statement of Theorem \ref{thm0}, we still need the following lemma.
%\begin{lem}\label{lemd4}
%Suppose $\mathcal{F}(y_1,\dots,y_m)=\prod_{i=1}^m(1-y_m)$ with $0\leq y_i\leq 1-\eta$ and $\sum_{i=1}^{m}y_i\geq a_\ast$, where $0<\eta<1$ and $a_\ast>0$ are two constants. Then we always have
%\begin{align}\label{91}
%\eta^m\leq\mathcal{F}(y_1,\dots,y_m)\leq \Big(1-\frac{a_\ast}{m}\Big)^m.
%\end{align}
%\end{lem}
%{\it Proof.} The lower bound part of (\ref{91}) is straightforward and the upper bound part follows the famous AM-GM inequality immediately. \hfill$\square$

We now  present the sufficiency proof of Theorem \ref{thm0}. In fact, we are going to prove a stronger statement which does not rely on the arc balance assumption A3.
\begin{prop}\label{prop3}
Suppose A1 and A2 hold. Global $\epsilon$-agreement is achieved for (\ref{9}) if $\mathcal{G}^p$ is quasi-strongly connected and there exist a constant $a_\ast>0$ and an integer $T_\ast>0$ such that $\sum_{s=t}^{t+T_\ast-1} W_{ij}(s)\geq a_\ast$ for all $t\geq0$ and $(j,i)\in \mathcal{G}^p$.
\end{prop}
{\it Proof.} Let $i_0\in\mathcal{V}$ be a center of $\mathcal{G}^p$. Take $t_0\geq0$. Assume first that
\begin{align}\label{d1}
x_{i_0}(t_0)\leq \frac{1}{2}\psi(t_0)+\frac{1}{2}\Psi(t_0).
 \end{align}
Then from Lemma \ref{lemd3}, one has
\begin{align}\label{92}
x_{i_0}(t_0+T)&\leq  \frac{1}{2}\prod_{s=t_0}^{t_0+T-1}\big(1- \xi^+(s;i_0)\big)\cdot\psi(t_0)+\Big(1-\frac{1}{2}\prod_{s=t_0}^{t_0+T-1}\big(1- \xi^+(s;i_0)\big)\Big)\cdot\Psi(t_0)\nonumber\\
&\leq \frac{\eta^{T}}{2} \psi(t_0)+\Big(1-\frac{\eta^{T}}{2}\Big)\Psi(t_0)
\end{align}
for all $T=0,1,\dots$.

 Denote $\mathcal{V}_1$ as the node set consisting of all the nodes of which $i_0$ is a neighbor in $\mathcal{G}^p$, i.e., $\mathcal{V}_1=\{j:(i_0,j)\in\mathcal{E}^p\}$. Note that $\mathcal{V}_1$ is nonempty because $i_0$ is a center. For any $i_1\in \mathcal{V}_1$, there exists an instance $\bar{t}_1\in[t_0,t_0+T_\ast-1]$ such that $W_{i_1i_0}(\bar{t}_1)\geq a_\ast/T_\ast$ because $\sum_{t=t_0}^{t_0+T_\ast-1}W_{i_1i_0}(t)\geq a_\ast$. Suppose $\bar{t}_1=t_0+\varrho_1$ with $\varrho_1 \in [0,T_\ast-1]$. Then with (\ref{92}), we have
 \begin{align}
x_{i_1}(\bar{t}_1+1)=x_{i_1}(t_0+\varrho_1+1)&\leq  W_{i_1i_0}(t_0+\varrho_1) x_{i_0}(t_0+\varrho_1)+\big(1-W_{i_1i_0}(t_0+\varrho_1)\big)\Psi(t_0)\nonumber\\
&\leq \frac{a_\ast}{T_\ast} \cdot \Big[ \frac{\eta^{\varrho_1}}{2} \psi(t_0)+\big(1-\frac{\eta^{\varrho_1}}{2}\big)\Psi(t_0) \Big]+\Big(1-\frac{a_\ast}{T_\ast}\Big)\Psi(t_0)\nonumber\\
&=\eta^{\varrho_1}\cdot \frac{a_\ast}{2T_\ast} \cdot \psi(t_0)+\Big(1-\eta^{\varrho_1}\cdot \frac{a_\ast}{2T_\ast}\Big)\Psi(t_0).
\end{align}
 Based on Lemma \ref{lemd3}, we can further conclude
  \begin{align}
x_{i_1}(t_0+\varrho_1+T)&\leq \eta^{\varrho_1+T-1}\cdot \frac{a_\ast}{2T_\ast} \cdot \psi(t_0)+\Big(1-\eta^{\varrho_1+T-1}\cdot \frac{a_\ast}{2T_\ast}\Big)\Psi(t_0)
\end{align}
for all $T=1,2,\dots$, which implies
  \begin{align}
x_{i_1}(t_0+T_\ast+K)&\leq \eta^{T_\ast+K}\cdot \frac{a_\ast}{2T_\ast} \cdot \psi(t_0)+\Big(1-\eta^{T_\ast+K}\cdot \frac{a_\ast}{2T_\ast}\Big)\Psi(t_0), \ \ K=0,1,\dots.
\end{align}

Next, since $\mathcal{G}^p$ is quasi-strongly connected, we can denote  $\mathcal{V}_2$ as the node set consisting of all the nodes each of which has a neighbor in $\{i_0\}\cup\mathcal{V}_1$ within  $\mathcal{G}^p$. For any $i_2\in \mathcal{V}_2$, there exist a node $i_\ast\in\{i_0\}\cup \mathcal{V}_1$ and an instance $\bar{t}_2=t_0+T_\ast+\varrho_2$ with $\varrho_2 \in [0,T_\ast-1]$ such that $W_{i_2i_\ast}(\bar{t}_1)\geq a_\ast/T_\ast$. Similarly we have
\begin{align}
x_{i_2}(\bar{t}_2+1)&\leq  W_{i_2i_\ast}(t_0+T_\ast+\varrho_2) x_{i_\ast}(t_0+T_\ast+\varrho_2)+\big(1-W_{i_2i_\ast}(t_0+T_\ast+\varrho_2)\big)\Psi(t_0)\nonumber\\
&\leq \frac{a_\ast}{T_\ast} \cdot \Big[ \eta^{T_\ast+\varrho_2}\cdot \frac{a_\ast}{2T_\ast} \cdot \psi(t_0)+\Big(1-\eta^{T_\ast+\varrho_2}\cdot \frac{a_\ast}{2T_\ast}\Big)\Psi(t_0) \Big]+\Big(1-\frac{a_\ast}{T_\ast}\Big)\Psi(t_0)\nonumber\\
&=\frac{\eta^{T_\ast+\varrho_2}}{2}\cdot \big(\frac{a_\ast}{T_\ast}\big)^2 \cdot \psi(t_0)+\Big(1-\frac{\eta^{T_\ast+\varrho_2}}{2}\cdot \big(\frac{a_\ast}{T_\ast}\big)^2\Big)\Psi(t_0),
\end{align}
and therefore
\begin{align}
x_{i_2}(t_0+2T_\ast+K)&\leq \frac{\eta^{2T_\ast+K}}{2}\cdot \big(\frac{a_\ast}{T_\ast}\big)^2 \psi(t_0)+\Big(1-\frac{\eta^{2T_\ast+K}}{2}\cdot \big(\frac{a_\ast}{T_\ast}\big)^2 \Big)\Psi(t_0), \ \ K=0,1,\dots \nonumber
\end{align}

 Proceeding the estimate, $\mathcal{V}_3,\dots,\mathcal{V}_{k}$ can be similarly defined until $\big(\cup_{i=1}^{k}\mathcal{V}_i\big)\cup\{i_0\}=\mathcal{V}$. Moreover, it is not hard to see that $i_0$ can be selected so that $k=d_0$, where $d_0$ is the diameter of $\mathcal{G}^p$, and thus
\begin{align}
x_{i}(t_0+d_0T_\ast)\leq \frac{\eta^{d_0T_\ast}}{2}\cdot \big(\frac{a_\ast}{T_\ast}\big)^{d_0} \cdot \psi(t_0)+\Big(1-\frac{\eta^{d_0T_\ast}}{2}\cdot \big(\frac{a_\ast}{T_\ast}\big)^{d_0}\Big)\Psi(t_0), \ \ i=1,\dots,n
\end{align}
which yields
\begin{align} \label{96}
\Psi(t_0+d_0T_\ast)\leq \frac{\eta^{d_0T_\ast}}{2}\cdot \big(\frac{a_\ast}{T_\ast}\big)^{d_0} \cdot \psi(t_0)+\Big(1-\frac{\eta^{d_0T_\ast}}{2}\cdot \big(\frac{a_\ast}{T_\ast}\big)^{d_0}\Big)\Psi(t_0).
\end{align}
With (\ref{96}), we eventually have
\begin{align} \label{99}
\mathcal{H}(t_0+d_0T_\ast)&\leq \frac{\eta^{d_0T_\ast}}{2}\cdot \big(\frac{a_\ast}{T_\ast}\big)^{d_0} \cdot \psi(t_0)+\Big(1-\frac{\eta^{d_0T_\ast}}{2}\cdot \big(\frac{a_\ast}{T_\ast}\big)^{d_0}\Big)\Psi(t_0)-\psi(t_0)\nonumber\\
&=\Big(1- \frac{\eta^{d_0T_\ast}}{2}\cdot \big(\frac{a_\ast}{T_\ast}\big)^{d_0}\Big) \mathcal{H}(t_0).
\end{align}

For the opposite case of (\ref{d1}) with
 \begin{align}
x_{i_0}(t_0)>\frac{1}{2}\psi(t_0)+\frac{1}{2}\Psi(t_0),
 \end{align}
(\ref{99}) is obtained using a symmetric argument by bounding $\psi(t_0+d_0T_\ast)$ from below.

Therefore, the desired conclusion follows with $\epsilon=1- \frac{\eta^{d_0T_\ast}}{2}\cdot \big(\frac{a_\ast}{T_\ast}\big)^2$ and $T_0=d_0T_\ast$ since (\ref{99}) holds independent with the choice of $t_0$. \hfill$\square$
%With assumption A1 and A2, we first notice that
%\begin{align}
%\frac{1- \xi^+(t;m)}{1- \xi_0^+(t;m)}&=\frac{1- \xi_0^+(t;m)- \xi_{\ast\setminus 0}^+(t;m)}{1- \xi_0^+(t;m)}=1-\frac{\xi_{\ast\setminus 0}^+(t;m)}{1- \xi_0^+(t;m)}\geq 1- %\frac{\xi_{\ast\setminus 0}^+(t;m)}{\eta+\xi_{\ast\setminus 0}^+(t;m)},
%\end{align}
%which implies
%\begin{align}\label{92}
%1- \xi^+(t;m) \geq \Big(1- \xi_0^+(t;m)\Big)\cdot\Big(1- \frac{\xi_{\ast\setminus 0}^+(t;m)}{\eta+\xi_{\ast\setminus 0}^+(t;m)}\Big).
%\end{align}

%According to the definition of persistent graph, we have $\sum_{t=0}^\infty{\xi_{\ast\setminus 0}^+(t;m)}<\infty$ thus also $\sum_{t=0}^\infty\frac{\xi_{\ast\setminus %0}^+(t;m)}{\eta+\xi_{\ast\setminus 0}^+(t;m)}<\infty$. As a result, based on Lemma \ref{lemd1}, we can well define
%$$
%\nu_0\doteq \min_{m\in\mathcal{V}} \ \ \prod_{t=0}^\infty \Big(1- \frac{\xi_{\ast\setminus 0}^+(t;m)}{\eta+\xi_{\ast\setminus 0}^+(t;m)}\Big)
%$$
%with $0<\nu_0\leq1$.

\section{Continuous-time Belief Evolution}
In this section, we turn to the continuous-time updating rule. We need an assumption on the continuity of each weight function $W_{ij}(t)$ for the existence of trajectories of (\ref{0}).

\vspace{2mm}

\noindent{\bf A4} {\it (Continuity)} Each $W_{ij}(t), \ (j,i)\in\mathcal{E}_\ast$ is continuous except for a set with measure zero.

With assumption A4, the set of discontinuity points for the right-hand side of equation (\ref{0}) has measure zero. Therefore,  the Caratheodory solutions of (\ref{0}) exist for arbitrary initial conditions, and they are absolutely continuous functions that satisfy (\ref{0}) for almost all $t$ on the maximum interval of existence  \cite{fili, cortes}. In the following, each solution of (\ref{0}) is considered in the sense of Caratheodory without explicit mention.

Let us first study the feasibility of the solutions of (\ref{0}). Consider (\ref{0}) with initial condition $x(t_0)=(x_1(t_0),\dots,x_n(t_0))^T=x^0\in \mathds{R}^n, t_0\geq0$.

The upper Dini
derivative of  a function $h: (a,b)\to \mathds{R}$ at $t$ is defined as
$$
D^+h(t)=\limsup_{s\to 0^+} \frac{h(t+s)-h(t)}{s}
$$
 The next result is useful for the calculation of Dini derivatives
 \cite{dan,lin07}.

\begin{lem}
\label{lem1}  Let $V_i(t,x): \mathds{R}\times \mathds{R}^m \to \mathds{R},\ i=1,\dots,n$, be
$C^1$ and $V(t,x)=\max_{i=1,\dots,n}V_i(t,x)$. If $
\mathcal{I}(t)=\big\{i\in \{1,\dots,n\}\,:\,V(t,x(t))=V_i(t,x(t))\big\}$
is the set of indices where the maximum is reached at $t$, then
$
D^+V(t,x(t))=\max_{i\in\mathcal{ I}(t)}\dot{V}_i(t,x(t)).
$
\end{lem}

The following lemma establishes the  monotonicity of $\Psi(t)$ and $\psi(t)$.
\begin{lem}
\label{lem2}For all $t\geq t_0\geq0$, we have $D^+\Psi(t)\leq 0$ and $D^+\psi(t)\geq 0.$
\end{lem}

\noindent{\it Proof.}  We prove $D^+\Psi(t)\leq 0$. The other part can be proved similarly.

Let $\mathcal{I}_0(t)$ represent the set containing all the agents
that reach the maximum in the definition of $\Psi(t)$ at
time $t$, i.e.,  $\mathcal{I}(t)=\{i\in\mathcal{V}|\ x_i(t)=\Psi(t)\}$. Then according to Lemma \ref{lem1}, we obtain
\begin{align}
D^+\Psi(t)=\max_{i\in\mathcal{I}_0(t)} \dot{x}_i(t)= \max_{i\in\mathcal{I}_0(t)} \Big[\sum\limits_{j \in
N_i}W_{ij}(t)(x_j(t)-x_i(t))\Big]\leq 0,
\end{align}
which completes the proof. \hfill$\square$

Lemma \ref{lem2} implies, $\mathcal{H}(t)$ is non-increasing for all $t\geq t_0$, and therefore each (Caratheodory) trajectory of (\ref{0}) is bounded within the initial states of the nodes. As a result, the trajectories exist in $[t_0,\infty)$ for any initial condition.

The main result on global consensus and  $\epsilon$-consensus is stated in the following two theorems.

\begin{thm}\label{thm1}
Suppose A3 and A4 hold. Global agreement is achieved for (\ref{0}) if and only if $\mathcal{G}^p$ is quasi-strongly connected.
\end{thm}

\begin{thm}\label{thm2}
Suppose A3 and A4 hold. Global $\epsilon$-agreement is achieved for (\ref{0}) if and only if

(a) $\mathcal{G}^p$ is quasi-strongly connected;

(b) there exists two constants $a_\ast$, $\tau_0>0$ such that $\int_{t}^{t+\tau_0} W_{ij}(s)ds\geq a_\ast$ for all $t\geq0$ and $(j,i)\in \mathcal{G}^p$.

 Moreover, if (a) and (b) hold, then we have
     \begin{align}
\mathcal{H}\Big(t+ \tau_0 \cdot \Big\lceil\frac{d_0\log 2}{a_\ast}\Big\rceil\Big)\leq  \Big(1-\frac{m_0^{d_0}}{2}\Big)\mathcal{H}(t),
\end{align}
where $m_0=\big(\frac{\omega_0}{2}\big)^2 \frac{1}{(n-1)A}$ with $\omega_0= e^{-\int_0^\infty\theta(t)dt}$, $d_0$ is the diameter of $\mathcal{G}^p$, and $\lceil z \rceil$ represents the smallest integer which is no smaller than $z$.
\end{thm}

Theorem \ref{thm1} implies that the connectivity of the persistent  graph $\mathcal{G}^p$ totally determines whether an agreement can be achieved globally. Furthermore, Theorem \ref{thm2} implies that $\int_0^T W_{ij}(t)dt=O(T)$ is a critical condition to ensure a global $\epsilon$-consensus.

\begin{remark}
Consensus for  (\ref{0}) was first studied in \cite{saber04}, where the convergence rate was shown to be determined by the second largest eigenvalue of the Laplacian of the communication graph.  Further discussions can be found in \cite{ren05,lin07,shi09}.
\end{remark}

\begin{remark}  Theorems \ref{thm1} and \ref{thm2} still hold if assumption A3 is replaced by the following integral version.

\noindent{\it A5.} {\it (Integral Arc Balance)} There exists a constant $A>1$ such that for any two arcs $(j,i), (m,k)\in\mathcal{E}^p$, we have
$$
A^{-1}\int_{a}^b W_{ij}(t) dt \leq \int_{a}^b W_{km}(t)dt \leq A\int_{a}^b W_{ij}(t)dt
$$
for all $0\leq a<b$.
\end{remark}

\begin{remark}
If we have $\int_{t=t_0}^T W_{ij}(t)dt=\infty, (j,i)\in\mathcal{G}^p$ for some finite $T$, it follows from the  proof of Theorem \ref{thm1} below  that (\ref{0}) will reach a global agreement in finite time when $t$ tends to $T$.
\end{remark}

\subsection{Preliminaries}

In this subsection,  we  establish two lemmas which describe the boundaries of how much each individual arc affects the nodes' dynamics. Then the proof of Theorems \ref{thm1} and \ref{thm2} will be proposed in the next two subsections.

\begin{lem}\label{lem3}
Suppose $x_{m}(s)\leq \mu\psi(s)+(1-\mu)\Psi(s)$ for some $s\geq t_0$ and $m\in\mathcal{V}$ with $0\leq\mu\leq1$ a giving constant. Then we have
\begin{align}
x_{m}(t)\leq \mu e^{-\int_{s}^t \xi^+(\tau;m)d\tau}\psi(s)+\big [1-\mu e^{-\int_{s}^t \xi^+(\tau;m)d\tau}\big] \Psi(s)
\end{align}
for all $t\geq s$.
\end{lem}
{\it Proof.} Based on Lemma \ref{lem2}, we see that
\begin{align}\label{10}
\dot{x}_{m}(t)= \sum_{j\in\mathcal{N}_m}W_{mj}(t)\big[x_j(t)-x_m(t)\big]
&\leq \sum_{j\in\mathcal{N}_m}W_{mj}(t)\big[\Psi(s)-x_m(t)\big]\nonumber\\
&= -\xi^+(t;m)\big[x_m(t)-\Psi(s)\big], \quad \ t\geq s.
\end{align}
This implies
\begin{align}
{x}_{m}(t)&\leq   e^{-\int_{s}^t \xi^+(\tau;m)d\tau}x_m(s)+\big [1-e^{-\int_{s}^t \xi^+(\tau;m)d\tau}\big ] \Psi(s)\nonumber\\
&\leq  \mu e^{-\int_{s}^t \xi^+(\tau;m)d\tau}\psi(s)+\big [1-\mu e^{-\int_{s}^t \xi^+(\tau;m)d\tau}\big ] \Psi(s)
\end{align}
by Gr\"{o}nwall's inequality. The proof is completed. \hfill$\square$

We give a lemma investigating the dynamic evolution between two connected nodes.

\begin{lem}\label{lem4}
Suppose $(l,m)\in \mathcal{E}_\ast$ and there exists a constant $0<\mu<1$ such that
  $$
  x_{l}(t)\leq \mu\psi(s_0)+(1-\mu)\Psi(s_0), \quad t\in[s_0,s]
  $$ for  $t_0\leq s_0<s$. Then we have
\begin{align}
x_{m}(t)&\leq \mu \int_{s_0} ^t e^{-\int_{u}^t \xi^+(\tau;m)d\tau}W_{ml}(u)du \cdot\psi(s_0)\nonumber\\
&\ \ \ \ \ \ \  +\Big[1-\mu \int_{s_0} ^t e^{-\int_{u}^t \xi^+(\tau;m)d\tau}W_{ml}(u)du \Big] \Psi(s_0), \quad t\in[s_0,s].
\end{align}
\end{lem}
{\it Proof.}  Similar to (\ref{10}), for any $t\in [s_0,s]$, we have
\begin{align}
\dot{x}_{m}(t)
\leq \big[\xi^+(t;m)-W_{ml}(t)\big]\cdot\big[\Psi(s_0)-x_m(t)\big] +W_{ml}(t)\big[\mu\psi(s_0)+(1-\mu)\Psi(s_0)-x_m(t)\big]. \nonumber
\end{align}

Therefore, noting the fact that
\begin{align}
 \int_{s_0} ^t e^{-\int_{u}^t \xi^+(\tau;m)d\tau}\xi^+(u;m)du
= \int_{s_0} ^t \frac{d}{du} \Big[e^{-\int_{u}^t \xi^+(\tau;m)d\tau}\Big]=1- e^{-\int_{s_0}^t \xi^+(\tau;m)d\tau},\nonumber
\end{align}
we obtain
\begin{align}
x_{m}(t)&\leq e^{-\int_{s_0}^t \xi^+(\tau;m)d\tau} x_{m}(s_0) + \int_{s_0} ^t e^{-\int_{u}^t \xi^+(\tau;m)d\tau} \big[\xi^+(u;m)-W_{ml}(u)\big]du \cdot \Psi(s_0)\nonumber\\
&\ \ \ \ \ \ \ +\int_{s_0} ^t e^{-\int_{u}^t \xi^+(\tau;m)d\tau}W_{ml}(u)du\cdot \Big[\mu\psi(s_0)+(1-\mu)\Psi(s_0)\Big]\nonumber\\
&\leq e^{-\int_{s_0}^t \xi^+(\tau;m)d\tau} \Psi(s_0) + \int_{s_0} ^t e^{-\int_{u}^t \xi^+(\tau;m)d\tau} \big[\xi^+(u;m)-W_{ml}(u)\big]du \cdot \Psi(s_0)\nonumber\\
&\ \ \ \ \ \ \ +\int_{s_0} ^t e^{-\int_{u}^t \xi^+(\tau;m)d\tau}W_{ml}(u)du\cdot \Big[\mu\psi(s_0)+(1-\mu)\Psi(s_0)\Big]\nonumber\\
&= \mu \int_{s_0} ^t e^{-\int_{u}^t \xi^+(\tau;m)d\tau}W_{ml}(u)du \cdot\psi(s_0)\nonumber\\
&\ \ \ \ \ \ \  +\Big[1-\mu \int_{s_0} ^t e^{-\int_{u}^t \xi^+(\tau;m)d\tau}W_{ml}(u)du \Big] \Psi(s_0),  \quad t\in[s_0,s]
\end{align}
by Gr\"{o}nwall's inequality and some simple manipulations. This completes the proof. \hfill$\square$
\subsection{Proof of Theorem \ref{thm1}}

\noindent{\it Sufficiency}

Let $i_0\in\mathcal{V}$ be a center of $\mathcal{G}^p$. Assume first that
\begin{align}\label{s1}
x_{i_0}(t_0)\leq \frac{1}{2}\psi(t_0)+\frac{1}{2}\Psi(t_0).
 \end{align}
Denote $\omega_0= e^{-\int_0^\infty\theta(t)dt}$. Then we have $0<\omega_0\leq 1$. Thus, based on Lemma \ref{lem3} and noting the fact that $\psi(t_0)\leq \Psi(t_0)$, we have
\begin{align}
{x}_{i_0}(t)&\leq  \frac{1}{2}e^{-\int_{t_0}^t \xi^+(\tau;i_0)d\tau}\psi(t_0)+\big[1-\frac{1}{2}e^{-\int_{t_0}^t \xi^+(\tau;i_0)d\tau}\big] \Psi(t_0)\nonumber\\
&\leq \frac{\omega_0 }{2}e^{-\int_{t_0}^t\xi_0^+(\tau;i_0)d\tau}\psi(t_0)+\Big[1-\frac{\omega_0 }{2} e^{-\int_{t_0}^t\xi_0^+(\tau;i_0)d\tau}\Big] \Psi(t_0).\nonumber
\end{align}

Define
\begin{align}\label{21}
\hat{t}_1= \inf \Big\{t\geq t_0:\  e^{- \int_{t_0}^t \xi_0^+(\tau;i_0)d\tau}=\frac{1}{2}  \Big\}.
\end{align}
We see that $\hat{t}_1$ is finite from the definition of $\mathcal{E}^p$. As a result, we obtain
\begin{align}
{x}_{i_0}(t)
\leq \frac{\omega_0}{4}\psi(t_0)+\big[1-\frac{\omega_0}{4}\big] \Psi(t_0),\ \ t\in[t_0,\hat{t}_1].
\end{align}

Next, we  denote the node set consisting of all the nodes of which $i_0$ is a neighbor in $\mathcal{G}^p$ as $\mathcal{V}_1$, i.e., $\mathcal{V}_1=\{j:(i_0,j)\in\mathcal{E}^p\}$. Note that $\mathcal{V}_1$ is nonempty because $i_0$ is a center. Then for any $i_1\in \mathcal{V}_1$, we see from Lemma \ref{lem4} that
\begin{align}\label{18}
x_{i_1}(\hat{t}_1)&\leq \frac{\omega_0}{4}  \int^{\hat{t}_1}_{t_0} e^{-\int_{u}^{\hat{t}_1} \xi^+(\tau;i_1)d\tau}W_{i_1i_0}(u)du \cdot\psi(t_0)\nonumber\\
&\ \ \ \ \ \ \  +\Big[1-\frac{\omega_0}{4}  \int^{\hat{t}_1} _{t_0} e^{-\int_{u}^{\hat{t}_1} \xi^+(\tau;i_1)d\tau}W_{i_1i_0}(u)du\Big] \Psi(s_0)\nonumber\\
&\leq \frac{\omega_0^2}{4}  \int^{\hat{t}_1} _{t_0} e^{-\int_{u}^{\hat{t}_1} \xi_0^+(\tau;i_1)d\tau}W_{i_1i_0}(u)du \cdot\psi(t_0)\nonumber\\
&\ \ \ \ \ \ \  +\Big[1-\frac{\omega_0^2}{4}  \int^{\hat{t}_1} _{t_0} e^{-\int_{u}^{\hat{t}_1} \xi_0^+(\tau;i_1)d\tau}W_{i_1i_0}(u)du\Big] \Psi(s_0).
\end{align}
The arc balance assumption A3 implies that
$$
\int_{u}^{\hat{t}_1}\xi_0^+(t;i_1)dt\leq \int_{u}^{\hat{t}_1}(n-1)A W_{i_1i_0}(t)dt,
$$
which yields
\begin{align}\label{16}
\int^{\hat{t}_1} _{t_0} e^{-\int_{u}^{\hat{t}_1} \xi_0^+(\tau;i_1)d\tau}W_{i_1i_0}(u)du &\geq \int^{\hat{t}_1} _{t_0} e^{-(n-1)A\int_{u}^{\hat{t}_1} W_{i_1i_0}(\tau)d\tau}W_{i_1i_0}(u)du\nonumber\\
&= \frac{1}{(n-1)A}\int^{\hat{t}_1} _{t_0} \frac{d}{du}e^{-(n-1)A\int_{u}^{\hat{t}_1} W_{i_1i_0}(\tau)d\tau}\nonumber\\
&=  \frac{1}{(n-1)A}\cdot \Big[1-e^{-(n-1)A\int_{t_0}^{\hat{t}_1} W_{i_1i_0}(\tau)d\tau}\Big].
\end{align}
On the other hand, we also have
$$
\int_{t_0}^{\hat{t}_1} \xi_0^+(t;i_0)dt \leq \int_{t_0}^{\hat{t}_1}(n-1)A W_{i_1i_0}(t)dt.
$$
Thus, we know from (\ref{16}) and the definition of $\hat{t}_1$  that
\begin{align}\label{17}
\int^{\hat{t}_1} _{t_0} e^{-\int_{u}^{\hat{t}_1} \xi_0^+(\tau;i_1)d\tau}W_{i_1i_0}(u)du &\geq  \frac{1}{(n-1)A}\cdot \Big[1-e^{-(n-1)A\int_{t_0}^{\hat{t}_1} W_{i_1i_0}(\tau)d\tau}\Big]\nonumber\\
&\geq \frac{1}{(n-1)A}\cdot \Big[1-e^{-\int_{t_0}^{\hat{t}_1} \xi_0^+(\tau;i_0)d\tau}\Big]\nonumber\\
&= \frac{1}{2(n-1)A}.
\end{align}
Equations (\ref{18}) and (\ref{17}) result in
\begin{align}\label{15}
x_{i_1}(\hat{t}_1)&\leq \frac{m_0}{2} \psi(t_0)+ (1-\frac{m_0}{2})\Psi(t_0)
\end{align}
for all $i_1\in\mathcal{V}_1$, where $m_0=\big(\frac{\omega_0}{2}\big)^2 \frac{1}{(n-1)A}$.

We continue to estimate the upper bound of nodes in $\{i_0\}\cup \mathcal{V}_1$ when $t\geq \hat{t}_1$. Define
$$
\mathcal{Y}(t)=\max_{i\in \{i_0\}\cup \mathcal{V}_1} x_i(t).
$$ Then $\mathcal{Y}(\hat{t}_1)\leq \frac{m_0}{2} \psi(t_0)+ (1-\frac{m_0}{2})\Psi(t_0)$. Similar to Lemma \ref{lem3}, we find that
$$
D^+ \mathcal{Y}(t)\leq -\beta(t)[\mathcal{Y}(t)-\Psi(\hat{t}_1)], \ \ t\geq \hat{t}_1,
$$
where $\beta(t)=\sum_{i\in \{i_0\}\cup \mathcal{V}_1, j\notin \{i_0\}\cup \mathcal{V}_1  }W_{ij}(t)$. This implies
\begin{align}
\mathcal{Y}(t)&\leq e^{-\int_{\hat{t}_1}^t \beta(\tau)d\tau}\mathcal{Y}(\hat{t}_1)+\Big[1-e^{-\int_{\hat{t}_1}^t \beta(\tau)d\tau}\Big]\Psi(\hat{t}_1)\nonumber\\
&\leq e^{-\int_{\hat{t}_1}^t \beta(\tau)d\tau}\Big[\frac{m_0}{2} \psi(t_0)+ (1-\frac{m_0}{2})\Psi(t_0)\Big]+\Big[1-e^{-\int_{\hat{t}_1}^t \beta(\tau)d\tau}\Big]\Psi({t}_0)\nonumber\\
&\leq \frac{m_0}{2} \cdot\omega_0 e^{-\int_{\hat{t}_1}^t \hat{\beta}(\tau)d\tau}\psi(t_0)+\Big[1- \frac{m_0}{2}\cdot \omega_0 e^{-\int_{\hat{t}_1}^t \hat{\beta}(\tau)d\tau}\Big]\Psi({t}_0),
\end{align}
where $\hat{\beta}(t)=\sum_{i\in \{i_0\}\cup \mathcal{V}_1, j\notin \{i_0\}\cup \mathcal{V}_1, (j,i)\in\mathcal{E}^p  }W_{ij}(t)$. We can then define
$$
\mathcal{V}_2=\Big\{j\notin \{i_0\}\cup \mathcal{V}_1: \exists i\in \{i_0\}\cup \mathcal{V}_1\ {\rm s.t.}\ (i,j)\in\mathcal{E}^p\Big\}
 $$
 and
 $$
 \hat{t}_2= \inf \Big\{t\geq \hat{t}_1:\  e^{- \int_{\hat{t}_1}^t \hat{\beta}(\tau)d\tau}=\frac{1}{2}  \Big\}
$$
and similar analysis with (\ref{15}) gives a bound to any  node $i_2\in \mathcal{V}_2$ as
\begin{align}\label{19}
x_{i_2}(\hat{t}_2)&\leq \frac{m_0^2}{2} \psi(t_0)+ \big(1-\frac{m_0^2}{2} \big)\Psi(t_0).
\end{align}
Moreover, (\ref{19}) also holds for nodes in $\{i_0\}\cup \mathcal{V}_1$.

Since $\mathcal{G}^p$ has a center, we can proceed the estimation to nodes in $\mathcal{V}_2,\dots, \mathcal{V}_k$ until $\big(\cup_{j=1}^k\mathcal{V}_j\big)\cup \{i_0\}=\mathcal{V}$ with $\hat{t}_{2},\dots, \hat{t}_{k}$ such that
\begin{align}
x_{i}(\hat{t}_{k})\leq \frac{m_0^k}{2} \psi(t_0)+ (1-\frac{m_0^k}{2})\Psi(t_0)
\end{align}
for all $i\in \mathcal{V}$,  which leads to
  \begin{align}
\Psi(\hat{t}_{k})\leq \frac{m_0^k}{2}\psi(t_0)+ (1-\frac{m_0^k}{2})\Psi(t_0).
\end{align}

We see that $i_0$ can be chosen so that $k\leq d_0$ always holds, where $d_0$ is the diameter of $\mathcal{G}^p$. Denoting $t_1=\hat t_{k}$, we eventually arrive at
    \begin{align}\label{22}
\mathcal{H}(t_1)=\Psi({t}_1)-\psi({t}_1)\leq \frac{m_0^{d_0}}{2} \psi(t_0)+ \big(1-\frac{m_0^{d_0}}{2} \big)\Psi(t_0)-\psi({t}_0)= \Big(1-\frac{m_0^{d_0}}{2} \Big)\mathcal{H}(t_0).
\end{align}

Although the analysis up to now is based on assumption (\ref{s1}), we see that (\ref{22}) also holds for the other case with $x_{i_0}(t_0)> \frac{1}{2}\psi(t_0)+\frac{1}{2}\Psi(t_0)$ using a symmetric argument by investigating the lower bound of $\psi(t_1)$.

Similar estimate can be carried out for $t_k,k=2,3,\dots$, which leads to
  \begin{align}
\mathcal{H}(t_{k+1})\leq  \Big(1-\frac{m_0^{d_0}}{2} \Big)\mathcal{H}(t_k)
\end{align}
for all $t_k, k=1,2,\dots$, which yields
\begin{align}
\mathcal{H}(t_{k})\leq \Big(1-\frac{m_0^{d_0}}{2} \Big)^k\mathcal{H}(t_0).
\end{align}
Therefore, we can now conclude that $\lim_{t\rightarrow \infty}\mathcal{H}(t)=0$ because $\mathcal{H}(t)$ is non-increasing and $0<m_0<1$. The sufficiency statement of Theorem \ref{thm1} is thus proved.

\vspace{2mm}

\noindent{\it Necessity}

We follow the same line as the proof of Proposition \ref{prop1}. Suppose $\mathcal{G}^p$ is not quasi-strongly connected. Let $\mathcal{V}_u$, $\mathcal{V}_w$, $\ell(t)$ and $\hbar(t)$ follow the definitions in the proof of Proposition \ref{prop1}. Also let $x_{i}(t_0)=0$ for all $i\in\mathcal{V}_u$, and $x_i(t_0)=1$ for all $i\in\mathcal{V}\setminus \mathcal{V}_u$.  According to Lemma \ref{lem2}, we  have $x_i(t)\in[0,1]$.

Based on Lemma \ref{lem1}, we have
\begin{align}\label{1}
D^+\ell(t)&=\max_{i\in\mathcal {I}_1(t)} \Big[\sum_{j\in\mathcal{N}_i} W_{ij}(t)\big(x_j(t)-x_i(t)\big)\Big]\nonumber\\
&\leq \max_{i\in\mathcal {I}_1(t)} \Big[\sum_{j\in\mathcal{N}_i\setminus \mathcal{V}_u} W_{ij}(t)\big(x_j(t)-x_i(t)\big)\Big]\nonumber\\
&= \max_{i\in\mathcal {I}_1(t)} \Big[\sum_{j\in\mathcal{N}_i,\ (j,i)\in \mathcal{E}_\ast\setminus \mathcal{E}^p } W_{ij}(t)\big(x_j(t)-x_i(t)\big)\Big]\nonumber\\
&\leq \theta(t)\cdot\big(1-\ell(t)\big)
\end{align}
where $\mathcal {I}_1(t)$ is the index set that contains the nodes where the maximum is reached and $\theta(t)$ is defined in (\ref{notation3}).

 Similarly we have
\begin{align}\label{2}
D^+\hbar(t)&= \min_{i\in\mathcal {I}_2(t)} \Big[\sum_{j\in\mathcal{N}_i} W_{ij}(t)\big(x_j(t)-x_i(t)\big)\Big]\nonumber\\
&\geq \min_{i\in\mathcal {I}_2(t)} \Big[\sum_{j\in\mathcal{N}_i\setminus \mathcal{V}_w} W_{ij}(t)\big(x_j(t)-x_i(t)\big)\Big]\nonumber\\
&= \min_{i\in\mathcal {I}_2(t)} \Big[\sum_{j\in\mathcal{N}_i,\ (j,i)\in \mathcal{E}_\ast\setminus \mathcal{E}^p } W_{ij}(t)\big(x_j(t)-x_i(t)\big)\Big]\nonumber\\
&\geq  -\theta(t)\cdot \hbar(t)
\end{align}
where $\mathcal {I}_2(t)$ is the index set that contains the nodes where the minimum is reached.

With (\ref{1}) and (\ref{2}), denoting $L(t)=\hbar(t)-\ell(t)$, we obtain
\begin{align}
D^+L(t)\geq  -\theta(t)\cdot (\hbar(t)-\ell(t)+1)=-\theta(t)\cdot (L(t)+1), \nonumber
\end{align}
which is equivalent to
\begin{align}\label{4}
D^+\Big[ e^{\int_{t_0}^t \theta(\tau)d\tau}(L(t)+1)\Big]\geq 0.
\end{align}
Therefore, we have
\begin{align}
L(t)&\geq  2e^{- \int_{t_0}^t \theta(\tau)d\tau}-1.
\end{align}

Since  $\int_{0}^\infty \theta(t) dt<\infty$, we can choose $t_0$ sufficiently large to ensure $e^{- \int_{t_0}^t \theta(\tau)d\tau}\geq\frac{2}{3}$
for all $t\geq t_0$. This leads to $\mathcal{H}(t)\geq L(t)\geq 1/3,\ t\geq t_0$. The necessity part of Theorem \ref{thm1} thus follows.

\subsection{Proof of Theorem \ref{thm2}}
We first prove the necessity statement. Based on Theorem \ref{thm1}, we only need to prove that  condition $(b)$ in Theorem \ref{thm2} is necessary. Suppose $(b)$ in Theorem \ref{thm2} does not hold. Then $\forall 0<\epsilon<1, T>0, \exists t_\ast(T,\epsilon)\geq 0$ and $(j_0,i_0)\in\mathcal{E}^p$ such that
\begin{align}\label{3}
\int_{t_\ast}^{t_\ast+T} W_{i_0j_0}(\tau)d\tau< A^{-1}(n-1)^{-1}\cdot \frac{\log\epsilon^{-1}}{2}.
\end{align}
 Since $(j_0,i_0)\in\mathcal{G}^p$, it is not hard to see that $t_\ast(T,\epsilon)\rightarrow \infty$ as $T\rightarrow\infty$ for any fixed $\epsilon$. Thus, without loss of generality, we can assume that (\ref{3}) also holds for the arcs in $\mathcal{E}_\ast \setminus\mathcal{E}^p$. Moreover, assumption A3 implies
\begin{align}\label{11}
\int_{t_\ast}^{t_\ast+T} W_{ij}(\tau)d\tau< (n-1)^{-1}\cdot \frac{\log\epsilon^{-1}}{2}
\end{align}
for all $(j,i)\in\mathcal{E}^p$.

From similar argument we used to obtain  (\ref{4}),
\begin{align}\label{6}
D^+\mathcal{H}(t)\geq  -2\Big[\sum_{ (j,i)\in  \mathcal{E}_\ast } W_{ij}(t)\Big]\mathcal{H}(t),\ \  t\geq t_0.
\end{align}
Therefore, letting the system initial time be $t_0=t_\ast$ with $\mathcal{H}(t_\ast)>0$, where $t_\ast$ is defined in (\ref{3}), we see from (\ref{3}) and (\ref{11}) that
\begin{align}\label{12}
2\sum_{ (j,i)\in  \mathcal{E}_\ast }\int_{t_\ast}^{t_\ast+T} W_{ij}(\tau)d\tau &< \log\epsilon^{-1}.
\end{align}
Consequently, (\ref{6}) and (\ref{12}) lead to
\begin{align}\label{7}
\mathcal{H}(t_\ast+T)&\geq e^{-2\sum_{ (j,i)\in  \mathcal{E}_\ast }\int_{t_\ast}^{t_\ast+T} W_{ij}(\tau)d\tau}\mathcal{H}(t_\ast)> \epsilon \mathcal{H}(t_\ast).
\end{align}
Then the necessity part of Theorem \ref{thm2} holds because $\epsilon$ and $T$ are arbitrarily chosen in (\ref{7}).

Next, we prove the sufficiency part of Theorem \ref{thm2} based on the convergence analysis in the proof of Theorem \ref{thm1}.

When there exist two constants $a_\ast$, $\tau_0>0$ such that $\int_{t}^{t+\tau_0} W_{ij}(\tau)d\tau\geq a_\ast$ for all $t\geq0$ and $(j,i)\in \mathcal{G}^p$, we have
\begin{equation}\label{s100}
\int_{t}^{t+\tau_0} b_0(\tau)d\tau\geq a_\ast,\ t\geq0,
\end{equation}
where $b_0(t)=\min_{(j,i)\in \mathcal{G}^p} W_{ij}(t)$.

Let us revisit the proof of Theorem \ref{thm1}. The definition of $\hat{t}_1$ in (\ref{21}) satisfies
\begin{align}
\hat{t}_1 =\inf \Big\{t\geq t_0:\  e^{- \int_{t_0}^t \xi_0^+(\tau;i_0)d\tau}=\frac{1}{2}  \Big\}\leq \inf \Big\{t\geq t_0:\ \  e^{-\int_{t_0}^tb_0(\tau)d\tau}=\frac{1}{2}\Big\}.
\end{align}
Similarly, for $\hat{t}_j,j=2,\dots,k$ with $k\leq d_0$, we have
\begin{align}
\hat{t}_{j} \leq \inf \Big\{t\geq \hat{t}_{j}:\ \  e^{-\int_{\hat{t}_{j-1}}^tb_0(\tau)d\tau}=\frac{1}{2}\Big\}.
\end{align}

Thus,  for $t_1=\hat{t}_k$ in (\ref{22}), it holds that
\begin{align}
{t}_1 \leq \inf \Big\{t\geq t_0:\ \  e^{-\int_{t_0}^tb_0(\tau)d\tau}=\Big(\frac{1}{2}\Big)^{d_0}\Big\}=\inf \Big\{t\geq t_0:\ \  \int_{t_0}^tb_0(\tau)d\tau=d_0\log 2\Big\}.
\end{align}
Based on (\ref{s100}), we have
$$
\Big\lfloor \frac{t-t_0}{\tau_0}\Big\rfloor a_\ast\leq \int_{t_0}^tb_0(\tau)d\tau,
$$
where $\lfloor z \rfloor$ represents the largest integer which is no larger than $z$. This immediately implies
\begin{align}
{t}_1 \leq t_0+\tau_0 \cdot \Big\lceil\frac{d_0\log 2}{a_\ast}\Big\rceil,
\end{align}
where $\lceil z \rceil$ represents the smallest integer which is no smaller than $z$.

Therefore, it can be concluded from Lemma \ref{lem2} and (\ref{22}) that
    \begin{align}\label{24}
\mathcal{H}\Big(t_0+ \tau_0 \cdot \Big\lceil\frac{d_0\log 2}{a_\ast}\Big\rceil\Big)\leq  \Big(1-\frac{m_0^{d_0}}{2}\Big)\mathcal{H}(t_0).
\end{align}
 The desired conclusion follows since (\ref{24}) holds independent with the choice of $t_0$. Thus, we have now completed the proof of Theorem \ref{thm2}.

\section{Discussions}
In this section, we  present some  comparisons between our results with existing work, and comparisons between the discrete-time and continuous-time belief evolutions.
\subsection{Relation to Cut-balanced Graphs}
In \cite{julien}, a cut-balance condition is introduced in the sense that there exists a constant $K\geq1$ such that for all $t$ and any nonempty subset $S\subseteq \mathcal{V}$, it holds that
\begin{align}\label{cut}
K^{-1}\sum_{i\in S,j\notin S}W_{ji}(t)\leq \sum_{i\in S,j\notin S}W_{ij}(t)\leq K\sum_{i\in S,j\notin S}W_{ji}(t).
\end{align}

If the persistent graph $\mathcal{G}^p$ is strongly connected, the arc balance assumption A3  implies condition (\ref{cut}) over $\mathcal{G}^p$. Therefore, in this particular case, assumption A3  is a special case of the cut-balance condition in \cite{julien}, though assumption (\ref{cut}) in \cite{julien} is  over the underlying graph $\mathcal{G}_\ast$ . Except for this slight difference, the convergence statements in Theorem \ref{thm0} and Theorem \ref{thm1} are consistent with the results given in \cite{julien} for strongly connected graphs.

On the other hand, when $\mathcal{G}^p$ is  quasi-strongly connected, the cut-balance condition never holds even under assumption A3, because there may be no arc pointing to the center node. Hence, in general, the results given in this paper  provides conditions for node agreement independent of the conditions in \cite{julien}.

\subsection{Discrete-time vs. Continuous-time}
Theorems \ref{thm0} and \ref{thm2} share quite similar structure and statement. However,  there are some essential differences between them.
\begin{itemize}
\item[(a)] The discrete-time result in Theorem \ref{thm0} highly relies on the self-confidence condition A2. Without A2, oscillations among the nodes may become inevitable and periodic solutions of (\ref{9}) may arise for almost all initial condition even under A1 and A3. Note that the arc balance condition A3 is only useful for the necessity part of Theorem \ref{thm0}.

    \item[(b)] For the continuous-time result in Theorem \ref{thm2}, each self weight $W_{ii}(t)$ does not even show up in the model (\ref{0}). The arc balance condition A3 is essential for the dynamics. Without A3, oscillations may occur if the arc weights of the persistent graph alternatively  become large.
\end{itemize}

Therefore, we can conclude that the self-confidence condition  is critical for discrete-time belief agreement, as is the arc balance condition for continuous-time case.

An interesting question is whether a similar conclusion can be made for the discrete-time model (\ref{9}) as the statement in Theorem \ref{thm1}. This question is open and  needs additional explorations. More general discussion on this problem can be found in \cite{touri} on the ergodicity of stochastic chains.

\section{Conclusions}
Individuals are equipped with beliefs in social activities. The evolution of the beliefs can be modeled as dynamical systems over graphs using for instance the widely studied consensus algorithms.  This paper studied persistent graphs under discrete-time and continuous-time consensus algorithms. Sufficient and necessary conditions were established on the persistent graph for the network to reach global agreement or $\epsilon$-agreement. It was shown that the persistent graph essentially determines both the convergence and convergence rate to an agreement.

\vspace{15mm}

\begin{center}
{\Large Acknowledgment }
\end{center}
The authors would like to thank Prof. Julien Hendrickx, Universit\'{e} Catholique de Louvain, for many helpful discussions.

\end{document}